\documentclass[a4paper,10pt,twocolumn,superscriptaddress,showpacs,aps,pra]{revtex4-1}
\usepackage{color}
\usepackage{newlfont}
\usepackage{graphicx}
\usepackage{amssymb}
\usepackage{amsmath}
\usepackage[latin1]{inputenc}

\newcommand{\ket}[1]{\mbox{$ | #1 \rangle $}}
\newcommand{\bra}[1]{\mbox{$ \langle #1 | $}}
\newcommand{\be}{\begin{eqnarray}}
\newcommand{\ee}{\end{eqnarray}}
\newcommand{\hp}{\textrm{\tiny $\mathrm{H}$}}

\begin{document}

\title{Integrability in time-dependent systems with one degree of freedom}
\author{R.~M.~Angelo}
\affiliation{Departamento de Física, Universidade Federal do Paraná, Caixa Postal 19044, 81531-990, Curitiba, PR, Brazil}
\author{E.~I.~Duzzioni}
\affiliation{Instituto de Física, Universidade Federal de Uberlândia, Caixa Postal 593, 38400-902, Uberlândia, MG, Brazil}
\author{A.~D.~Ribeiro}
\affiliation{Departamento de Física, Universidade Federal do Paraná, Caixa Postal 19044, 81531-990, Curitiba, PR, Brazil}

\begin{abstract}
The notion of integrability is discussed for classical nonautonomous systems with one degree of freedom. The analysis is focused on models which are linearly spanned by finite Lie algebras. By constructing the autonomous extension of the time-dependent Hamiltonian we prove the existence of two invariants in involution which are shown to obey the criterion of functional independence. The implication of this result is that chaotic motion cannot exist in these systems. In addition, if the invariant manifold is compact, then the system is Liouville integrable. As an application, we discuss regimes of integrability in models of dynamical tunneling and parametric resonance, and in the dynamics of two-level systems under generic classical fields. A corresponding quantum algebraic structure is shown to exist which satisfies analog conditions of Liouville integrability and reproduces the classical dynamics in an appropriate limit within the Weyl-Wigner formalism. The quantum analog is then conjectured to be integrable as well.
\end{abstract}



\maketitle

\section{Introduction}

The importance of the notion of integrability in classical mechanics is unquestionable. This is so because it allows for the whole set of dynamical systems to be classified in two well distinguishable categories, namely, the regular (integrable) and chaotic (nonintengrable) systems. In particular, this discrimination finds rigorous support in the Arnold-Liouville theorem~\cite{gutz,ozorio,arnold}, which guarantees distinct properties for the dynamics of systems with a sufficient number of invariants. Among these properties a noticeable one is the confinement of the dynamical flux to invariant tori. This implies that angle-action variables exist and that the canonical equations of motion can be integrated by quadratures~\cite{arnold}. Furthermore, the Arnold-Liouville theorem provides insights for the definition of operational tools, such as Lyapunov exponents and Poincaré maps, which allow for the classification of the system even when the conditions of the theorem are not found. 

The above mentioned theory of integrability is elaborated for autonomous systems, whose Hamiltonian is a constant of motion. However, there is an unquotable number of works, in many different fields of physics and mathematics, dedicated to systems describable by nonautonomous Hamiltonians, which depend explicitly on time and hence are no longer time-invariant quantities (see, e.g., Refs.~\cite{pomeau,hensinger,soto10} and references therein for a few examples). Within this category, we find a very important subset composed of systems with {\em 1.5 degrees of freedom}, whose formulation is given in terms of a Hamiltonian $H(x,t)$ with one degree of freedom, $x=(q,p)$, plus explicit time dependence. These models have shown to be of great relevance for many theoretical and experimental investigations, as for instance in the problem of dynamical tunneling~\cite{hensinger,shudo}. In virtue of their time dependence, Hamiltonians of type $H(x,t)$ are known to display a rather rich dynamics, chaotic in general, kicked systems being emblematic examples~\cite{shudo}. In some cases, on the other hand, although complex, the dynamics is still regular, as in the case of the polemic {\em chaotic Rabi oscillations} of two-level systems under quasiperiodic classical fields~\cite{pomeau,badii,angelo05}. This model will be assessed here in the light of our results.

In the context of nonautonomous formulations, {\em Lie Hamiltonian systems} are particularly important. Because they are, by definition, linearly spanned by a finite Lie algebra, several mathematical results have succeeded to be derived which assess the conditions for the explicit integrability of the equations of motion, in both classical and quantum mechanics~\cite{jfc00,jfc08_jpa,jfc08_pla,dis,rfe_11,kn_10}. This paper is devoted to discuss the integrability in this arena. In the first part, we show that models with one degree of freedom are integrable. Our result establishes sufficient conditions for the absence of chaos in this important class of systems and hence gives rigorous arguments attesting the integrability of the aforementioned two-level systems.

Next, we proceed to the extension of our result to the quantum domain. Attempts on the elaboration of the notion of quantum integrability date back to the 1980's with Korsch~\cite{korsch82}, who claimed that a classically integrable system would naturally be quantumly integrable as well. Soon after, however, Hietarinta~\cite{hietarinta82,hietarinta84} showed that Korsch's rules for the quantum-classical correspondence are not generally applicable. Later on, Zhang and co-authors~\cite{zhang89} discussed the problem of the ambiguity underlying the concept of quantum degrees of freedom and proposed a definition of quantum integrability in terms of dynamical symmetries. Further, Weigert~\cite{weigert92} and Gallardo and co-authors~\cite{gallardo09} highlighted difficulties in translating the concept of {\em independent} constants to the quantum formalism. In addition, it is possible to quote many other different approaches to the problem of quantum integrability (see Refs.~\cite{zhang89,zhang90,caux10,weid09,gorin06} and references therein). Despite this long-standing debate on the notion of quantum integrability there is no consensual definition to date.

Our contribution will consist in constructing a quantum model in one-to-one correspondence with the classical one, specifically in what regards the Lie algebraic structure and the existence of a dynamical invariant. We will show that the corresponding quantum system satisfies conditions analogous to those which preclude the existence of chaos in the classical model. As a consequence, we will naturally be lead to conjecture in favor of the integrability of the quantum analog.

In order to prove our results we will gather some important methods often applied in different contexts. The theory of dynamical invariants, which was born long ago with the seminal works by Noether~\cite{noether}, Lutzky~\cite{lutzky}, and Lewis~\cite{lewis}, and has gained increasing interest through the consecutive decades~\cite{korsch79,kaushal81,kaushal93,bouquet98,riedel01,vazquez08}, will be invoked to search for invariants. Another valuable tool, usually subsidiary to the theory of invariants, is the notion of dynamical algebra~\cite{korsch79,kaushal81,kaushal93}. The conjugation of these two methods has proven to be of great relevance to the approach of a variety of quantum problems~\cite{yeh93,nunez94,bandy01,shen03,lima08,soto10,suslov10}, after pioneer studies by Lewis and Riesenfeld~\cite{lewis,lr}. In addition, we will apply Howland's method~\cite{howland,jauslin,lich} to construct an autonomous extension for the time-dependent Hamiltonian and, finally, the Weyl-Wigner formalism~\cite{weyl27,wigner32,polko10,moyal49} to establish the quantum-classical link.

\section{Preliminary notions}

In this paper we often invoke the notions of integrability and chaos as two complementary properties. For completeness, we review some commonly used statements about these concepts.

Although there is no universal definition of chaos, some conditions are usually associated to the existence of chaotic motion in dynamical systems. In this sense, one may regard them as a definition of chaos~\cite{devaney,hasselblatt}.

\vskip3mm
\noindent
{\bf Definition 1 (Devaney's chaos).} {\em A map $f: X\to~X $ is said to be chaotic on a space $X$ if (i) $f$ has sensitive dependence on initial conditions, (ii) $f$ is topologically transitive, and (iii) the periodic points are dense in $X$.}
\vskip3mm
Actually, it has been recognized~\cite{hasselblatt} that the last two (topological) conditions in fact imply sensitivity to initial conditions, so that the first condition above, which is far more relevant for practical purposes, turns out to be redundant. 

Topological transitivity (or topological mixing~\cite{hasselblatt}), on the other hand, is crucial. $f$ is said to be {\em transitive} if for any two nonempty open subsets $U$ and $V$ of $X$, there is a natural number $n$ such that $f^n(U)\cap V\neq\emptyset$. Intuitively, it means that any given region or set of points in $X$ will eventually overlap with any other given region. Equivalently, transitivity also means that $f$ has a dense orbit~\cite{hasselblatt}. By definition, one expects transitivity to exist only when the motion occurs on {\em compact} and {\em connected} manifolds. That is why the flows generated by some simple one-dimensional Hamiltonians, such as $H= q p$ or $H=p^2-q^2$, though exponentially sensitive to initial conditions, are not chaotic: the energy surfaces are not compact and the map cannot be transitive.

Density of periodic points means that every point in space is arbitrarily closely approached by periodic orbits. The conjugation of topological transitivity with periodic orbits implies the global orbit structure to be marked by an inextricable intertwining of density and periodicity. This leads to {\em homoclinic} and {\em heteroclinic tangle}~\cite{jose} and hence to sensitive dependence on initial conditions.

It is also common to define chaos in terms of its negation. This is motivated by the Arnold-Liouville theorem~\cite{arnold}, which defines the class of {\em completely integrable systems} (or Liouville-integrable systems).

\vskip3mm
\noindent
{\bf Theorem 1 (Arnold-Liouville).} {\em Suppose we are given $n$ functions $F_1,\cdots,F_n$ in involution,  $\{F_i,F_j\}=0$, and functionally independent on a level set 
\be
M_{\mathbf{f}}=\{x:F_i(x)=f_i,i=1,\cdots,n\}\nonumber
\ee 
on the $2n$-dimensional manifold. Then (i) $M_{\mathbf{f}}$ is a smooth $n$-dimensional manifold, invariant under the phase flow with Hamiltonian function $H=F_1$. In addition, if the manifold $M_{\mathbf{f}}$ is compact and connected, then (ii) it is diffeomorphic to a $n$-dimensional torus, the phase flow generated by $H$ is conditionally periodic (so that angle-action variables do exist in a neighborhood of $M_{\mathbf{f}}$), and the canonical equations can be integrated by quadratures.}
\vskip3mm

Liouville integrability is, therefore, synonymous of absence of chaos: the motion is confined to invariant submanifolds, so that topological transitivity does not exist on the energetically accessible space, the motion is fundamentally periodic and the solutions of the equations of motion are trivially integrated by quadratures. Thus, there is no room for complex motion. 

It is important to note at this point that the theorem ensures integrability only when the invariant manifold is compact and connected. This condition is commonly ignored but it is in fact not automatically implied by the existence of the invariants. Moreover, the theorem does not provide any mathematical test allowing for the discrimination of the manifold features.

\section{Classical integrability}

We are now ready to state and prove our first result. It establishes {\em sufficient conditions} for Lie systems with 1.5 degrees of freedom to be integrable. Differently from the Lie-Scheffers theorem (see Theorem 1 in Ref.~\cite{jfc00}), which guarantees the existence of a superposition rule for Lie systems with arbitrary dimension but requires a set of particular solutions to be found in order to construct the explicit solution, here we ensure integrability (absence of chaos), but for the dimension one and a half.

\vskip3mm
\noindent
{\bf Proposition 1.} {\em Let $H(x,t)$ be the Hamiltonian of a nonautonomous system with a single degree of freedom, $x=(q,p)$, where $q$ and $p$ are canonical variables satisfying $\{q,p\}=1$. If the Hamiltonian is expansible as a linear combination of the elements $\{O_1,O_2,\cdots,O_M\}$ of a finite Lie algebra, i.e.,
\begin{subequations}
\be
H(x,t)=\mathbf{h}(t)\cdot\mathbf{O}(x)=\sum\limits_{i=1}^{M}h_i(t)\,O_i(x)
\label{HO}
\ee
and
\be
\{O_i,O_j\}=\sum\limits_{k=1}^M\gamma_{ij}^k\,O_k,
\label{Ocommut}
\ee
where $\gamma_{ij}^k=-\gamma_{ji}^k$, and if $h_i(t)$ are complex-valued functions continuous on the interval $a\leqslant t\leqslant b$, then (i) an invariant of form
\be
I(x,t)=\mathbf{g}(t)\cdot\mathbf{O}(x)=\sum\limits_{i=1}^Mg_i(t)\,O_i(x)
\label{IO}
\ee
exists on the interval. Let $K(x,y)$ and $I(x,\theta)$, where $y=(\theta,J)$, be the autonomous extensions of the Hamiltonian and the invariant, respectively. If the two-dimensional manifold 
\be
M_{\mathcal{KI}}=\{(x,y):K(x,y)=\mathcal{K},\,I(x,\theta)=\mathcal{I}\}
\label{M_KI}
\ee
\end{subequations}
is compact and connected, then (ii) $H(x,t)$ is Liouville integrable. If $M_{\mathcal{KI}}$ is otherwise noncompact, then (iii) the dynamics generated by $H(x,t)$ is nonchaotic.}
\vskip3mm

The remainder of this section is devoted to the proof of this proposition and its application to systems of physical interest. First, we invoke the Lewis-Reisenfeld theory~\cite{lr}, according to which an invariant of form \eqref{IO} is required to satisfy the relation
\be
\partial_t I(x,t)+\left\{ I(x,t),H(x,t)\right\}=0,
\label{dI/dt}
\ee
so as to ensure that $\dot{I}=0$. By plugging Eqs.~\eqref{HO} and \eqref{IO} into Eq.~\eqref{dI/dt} one finds that
\be
\dot{g}_{k}(t)=\sum_{i,j=1}^{M}h_{i}(t)\,\gamma_{ij}^k\,g_{j}(t). 
\label{dg/dt}
\ee
This linear system of coupled differential equations can be summarized in matrix form as $\dot{\mathbf{g}}=\mathbf{F}(t)\mathbf{g}$. Now, if $\mathbf{h}(t)$ is continuous within the interval $a\leqslant t\leqslant b$, if $a\leqslant t_0\leqslant b$, and if $|\mathbf{g}_0|<\infty$, then the theorem of existence and uniqueness guarantees that the above system has a unique solution $\mathbf{g}(t)$ with initial condition $\mathbf{g}(t_0)=\mathbf{g}_0$ existing on the interval (see, e.g., Ref.~\cite{hartman}). It follows that a dynamical invariant of form \eqref{IO}, whose initial value $I(x_0,0)$ is not fixed a priori, will exist on the interval, thus proving part (i) of the proposition. Note that for most physical systems the interval of existence comprises the entire positive real line. Also, it is worth mentioning that this result consists in a generalization of those recently reported in Ref.~\cite{rfe_11}.

Theorem 1 requires, as a first condition to diagnose integrability, as many invariants as the number of degrees of freedom. Here we have proved the existence of a single invariant for a system with 1.5 degrees of freedom. Then the question arises whether it is enough. As we show now the answer is positive. Our strategy consists in moving to a two-dimensional formulation and then proving the existence of two functionally independent constants in involution.  

We start by constructing the autonomous two-dimensional extension of $H(x,t)$. Our problem of initial value can be written as
\be
\begin{array}{l}
\displaystyle{\frac{d x}{dt}=\{x,H\}}, \\ \\
x(0)=x_0,
\end{array} \label{HEH}
\ee
where $x_0=(q_0,p_0)$. The procedure consists in introducing an {\em extended} dynamical problem,
\be
\begin{array}{l}
\displaystyle{\frac{d x}{dt}=\{x,K\}},\qquad
\displaystyle{\frac{d y}{dt}=\{y,K\}},\\ \\
(x(0),y(0))=(x_0,y_0),
\end{array}\label{HEK}
\ee
governed by an {\em autonomous} Hamiltonian
\be
\begin{array}{l}
K(x,y):= H(x,\theta)+J,
\end{array}\label{K=H+J}
\ee
with an additional degree of freedom, $y=(\theta,J)$, such that $y_0=(0,J_0)$ and $\{\theta,J\}=1$. In effect, the equivalence with the original dynamics can be straightforwardly verified by realizing, from Hamilton's equation, that $\theta(t)=t$. It is then immediate that Eq.~\eqref{HEK} reproduces Eq.~\eqref{HEH}. This scheme is sometimes referred to as Howland's method~\cite{howland,jauslin}. It takes us to an autonomous formulation within which the new Hamiltonian $K$ is a constant of motion, a fact than can be easily proved from the Hamilton equation for $J$. 

Under the replacement $t\to\theta$, one then shows that $I(x,t)$ is an invariant in involution with the autonomous Hamiltonian $K(x,y)$. Let us compute the time derivative of the invariant:
\be
\dot{I}=\{I,K\}&=&\{I,K\}_x+\{I,K\}_y \nonumber \\
&=&\{I,H\}_x+\partial_{\theta}I.
\ee
Since $\theta=t$ the above relation implies, by Eq.~\eqref{dI/dt}, that $\{I(x,\theta),K(x,y)\}=0$, as we wanted to show. Therefore, we have moved to a two-degrees-of-freedom description for which two invariants exist in involution.

We are now left with the task which is often neglected (as pointed out in Ref.~\cite{weigert92}) in many treatments of quantum and classical integrability: the assessment of the {\em functional independence}. In classical mechanics, this notion can be stated as follows~\cite{arnold,ozorio}. 

\vskip3mm\noindent
{\bf Definition 2.} {\em Let the flow direction produced by an invariant $I_k$ be defined by the phase-space velocity $\mathbf{v}_{I_k}=(d\mathbf{r}/dt)_{I_k}=\mathcal{L}I_k$, where $\mathcal{L}=\mathcal{L}_x+\mathcal{L}_y$, $\mathcal{L}_{x,y}I_k:= \{\mathbf{r},I_k\}_{x,y}$, and $\mathbf{r}=(x,y)$. The set of invariants $\{I_1,I_2,\cdots,I_N\}$ is functionally independent if the vectors $\mathbf{v}_{I_1},\mathbf{v}_{I_2},\cdots,\mathbf{v}_{I_N}$ are linearly independent.}
\vskip3mm

Let $\mathbf{V}=a_K\mathbf{v}_{K}+a_I\mathbf{v}_{I}$ be a linear combination of the phase-space velocities generated by $K$ and $I$. By noticing that $K(x,y)=\mathbf{h}(\theta)\cdot\mathbf{O}(x)+J$ and $I(x,\theta)=\mathbf{g}(\theta)\cdot\mathbf{O}(x)$, we write 
\be
\mathbf{V}&=&(\mathcal{L}_{x}+\mathcal{L}_{y})(a_K\,K+a_I\,I) \nonumber \\
&=&[a_K\,\mathbf{h}(\theta)+a_I\,\mathbf{g}(\theta)]\cdot\mathcal{L}_{x}\mathbf{O}(x)\nonumber \\ &-&\left[a_K\,\mathbf{h}'(\theta)+a_I\,\mathbf{g}'(\theta)\right]\cdot\mathbf{O}(x)\,\mathbf{e}_{\textrm{\tiny $J$}} +a_K\,\mathbf{e}_{\textrm{\tiny $\theta$}},
\label{V}
\ee
where $a_{K,I}$ are arbitrary real numbers and $\mathbf{e}_{\textrm{\tiny $\theta,J$}}$ are unitary vectors denoting orthogonal directions in phase space. 

The proof of the functional independence of $K$ on $I$ is given by {\em reductio ad absurdum}. Assume that $\mathbf{v}_{K}$ and $\mathbf{v}_{I}$ are linearly dependent, so that the solution for $\mathbf{V}=0$ is to be given for nonvanishing $a_K$ and $a_I$. However, from Eq.~\eqref{V} we see that the only solution allowed is $a_K=a_I=0$, which contradicts the initial assumption. Hence, $\mathbf{v}_{K}$ and $\mathbf{v}_{I}$ are linearly independent and $K(x,y)$ and $I(x,\theta)$ are functionally independent. 

It follows, by implication (i) of the Arnold-Liouville theorem, that the evolution of any initial condition is confined to a two-dimensional invariant manifold $M_{\mathcal{KI}}$, as defined by Eq.~\eqref{M_KI}. Then, if $M_{\mathcal{KI}}$ is compact and connected, Theorem 1 immediately ensures Liouville integrability for $K(x,y)$. Given the proven dynamical equivalence of $K(x,y)$ with its nonautonomous version $H(x,t)$ we concluded that the latter is Liouville integrable as well, thus proving part (ii) of the proposition.

On the other hand, if $M_{\mathcal{KI}}$ is noncompact (unlimited) the existence of recurrent orbits is no longer guaranteed by the Poincaré recurrence theorem. As a consequence, a given orbit cannot densely fulfill its energetically accessible manifold and hence the dynamics cannot be transitive in general. Then, since one of the conditions for Definition 1 is not satisfied the motion is not chaotic. This argument is better appreciated by the formal construction of the Poincaré map for the autonomous Hamiltonian $K(x,y)$.

Consider a map $f:X\to X$ constructed by the collection of points $x_n=(q_n,p_n)$ obtained from the flow at the instants $t_n$ for which the orbit crosses a plane, say, $J_n=J(t_n)=0$. The set of points allowed for the map necessarily belongs to the level set $M_{\mathcal{KI}}$ where the orbit lives on. For a fixed energy $E$, the choice of an initial condition $x_{\nu}$, enumerated in $X$ by the real subindex $\nu$, gives the initial value $\theta_{\nu}$ by the relation $K(x_{\nu},\theta_{\nu},0)=E$. Moreover, the value $E$ defines all points $x\in X$. It follows that the invariant assumes a specific value $I(x_{\nu},\theta_{\nu})=\mathcal{I}_{\nu}$ and thus each initial condition $(x_{\nu},\theta_{\nu},0)$ evolves in time on a manifold $M_{E\mathcal{I}_{\nu}}=\{(x,y):K(x,y)=E,I(x,\theta)=\mathcal{I}_{\nu})\}$. The intersection of this manifold with the plane $J_n=0$, for every $n$, can be denoted by
\be\begin{array}{l}
K_{\nu}(x_n,\theta_n,0):= K(x_{\nu},\theta_{\nu},0)=E, \\
I_{\nu}(x_n,\theta_n,0):= I(x_{\nu},\theta_{\nu},0)=\mathcal{I}_{\nu}.\label{M_EI_nu}
\end{array}\ee
The subindex $\nu$ has been added in $K_{\nu}(x_n,\theta_n,0)$ and $I_{\nu}(x_n,\theta_n,0)$ to discriminate the initial condition which the map refers to. Actually, $\nu$ may be thought of as denoting the {\em set} of initial conditions satisfying \eqref{M_EI_nu} for fixed values of $\mathcal{I}_{\nu}$ and $E$. By formally inverting the former of the above relations, so as to get $\theta_n=\theta_n(x_n,E)$, we rewrite the latter as a function $\mathcal{I}$ of $x_n$ and $E$ only:
\be
\mathcal{I}(x_n,E):=I_{\nu}(x_n,\theta_n(x_n,E),0)=\mathcal{I}_{\nu}.
\ee
The set of curves $\mathcal{I}_{\nu}$ is dense in $X$ and each single curve does not intersect another. Now, let us take the set
\be
U_{\bar{\nu},\delta}=\Big\{x_n:\mathcal{I}(x_n,E)=\mathcal{I}_{\nu};\nu\in\left[\bar{\nu}-\delta,\bar{\nu}+\delta\right]\Big\},
\ee
composed of the points $x_n$ of a contiguous set of curves $\mathcal{I}_{\nu}$ such that $\nu\in\left[\bar{\nu}-\delta,\bar{\nu}+\delta\right]$. Since each $\mathcal{I}_{\nu}$ is invariant under the the action of the map, it follows that $f^m\left(U_{\bar{\nu},\delta}\right)=U_{\bar{\nu},\delta}$ for any natural number $m$. Consider another subset, e.g., $V=U_{\bar{\nu}+3\delta,\delta}$. Clearly, $f^{m}\left(U_{\bar{\nu},\delta}\right)\cap V=\emptyset$, so that the map $f$ is not topologically transitive and hence nonchaotic, no matter whether the manifold $M_{E\mathcal{I}_{\nu}}$ is compact or not. This completes the proof of our result. \hfill{$\blacksquare$}

It is worth mentioning that in cases in which $I(x,t)$ is explicitly known one can invoke a result by Bouquet and Bourdier~\cite{bouquet98} which shows how to construct a second invariant and then integrate the canonical equations by quadratures, i.e., by using only algebraic operations (including the inverse of functions) and integrals of known functions. This formal result implies that the Hamiltonian \eqref{HO} is, in principle, integrable, though not necessarily Liouville integrable. It follows also from this analysis that there is no chaos, in accordance with our result.

\subsection{Applications}

To illustrate the relevance of our result we employ it to assess some important physical problems. 
First, we provide some numerical results for a model of {\em dynamical tunneling}, phenomenon which has been experimentally investigated with ultracold atoms in Bose-Einstein condensates~\cite{hensinger}. The Hamiltonian may be written as
\be
H(x,t)=\frac{p^2}{2}+\Omega(t)\frac{q^2}{2}+\epsilon\frac{q^4}{4},
\label{H_casestudy}
\ee
where $\Omega(t)$ is a given function of time. (In what follows, numerical values of all pertinent physical quantities, including $\Omega(t)$ and $\epsilon$, will be given in arbitrary units.)

\begin{figure}
\centerline{\includegraphics[scale=0.6]{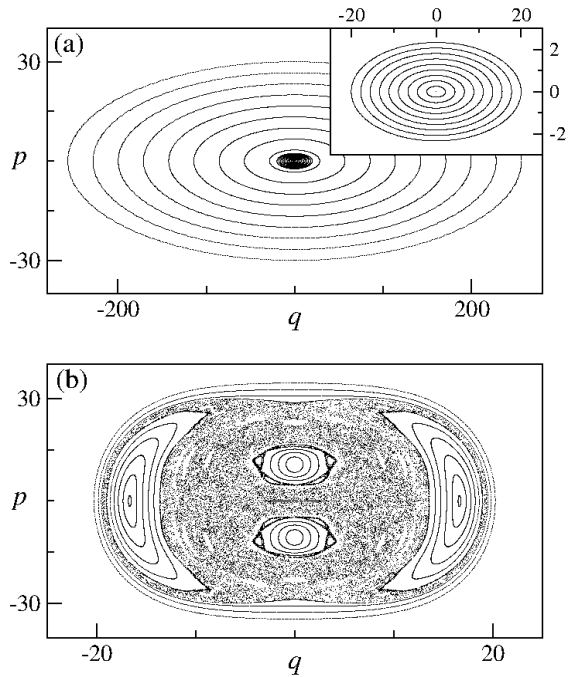}}
\caption{Poincaré map, in arbitrary units, for the periodic function $\Omega(t)=\cos(\pi t/2)$. In (a) $\epsilon=0$ and in (b) $\epsilon=0.01$. The inset in (a) corresponds to a zoom in the central dark region. The introduction of the term $\epsilon\,q^4/4$, even perturbativelly, opens the Lie algebra, thus breaking the sufficient conditions for Liouville integrability. This fact is denounced by the appearance of the chaotic sea in (b).}
\label{fig1}
\end{figure}

If $\epsilon=0$ the Hamiltonian reduces to the form \eqref{HO}, with the algebra $\{1,q,p,qp,q^2,p^2\}$. It follows from our result that the system is integrable for any well-behaved $\Omega(t)$. On the other hand, for any nonvanishing $\epsilon$, even if arbitrarily small, the algebra no longer closes and one cannot ensure integrability anymore. In Fig.~\ref{fig1} a numerical simulation is presented which illustrates the breakdown of the integrability when the quartic term is switched on for a dynamics under the function 
\be
\Omega(t)=\cos(\pi t/2).
\ee 
Since $\Omega (t)$ is periodic, the Poincaré section turns out to be equivalent to a {\em stroboscopic map}, which is constructed by collecting points $x_n=(q(t_n),p(t_n))$ such that $t_n=2\pi n/\omega$. In this example, $\omega=\pi/2$. All numerical simulations in this paper were performed with a Runge-Kutta integrator of 5th order.

Notably, this is a situation in which the invariant is compact and connected, as can be inferred by the tori structure in Fig.~\ref{fig1}-(a). Hence, the system is Liouville integrable. When the quartic term $\epsilon\,q^4/4$ is introduced, many rational tori are destroyed, the Poincaré map becomes locally transitive, and sensitivity to initial conditions starts to exist. In Fig.~\eqref{fig1}-(b) we see some of the surviving KAM-tori within the chaotic sea.

\begin{figure}
\centerline{\includegraphics[scale=0.6]{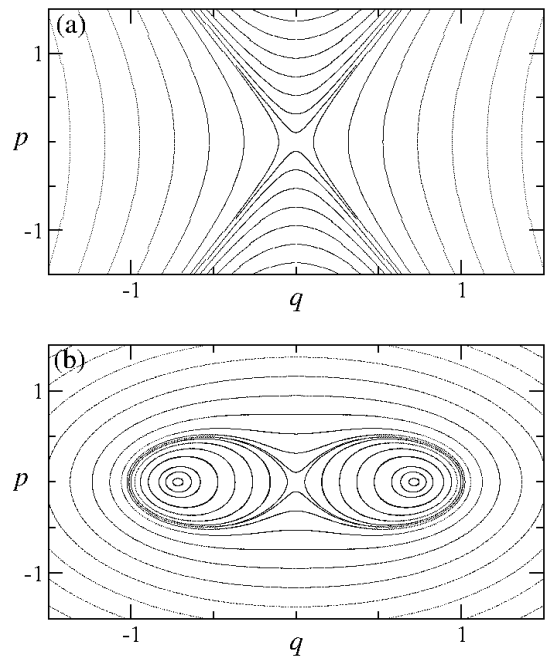}}
\caption{Poincaré map, in arbitrary units, for the resonant function $\Omega(t)=1+\textrm{\scriptsize $\frac{1}{10}$}\cos t$. In (a) $\epsilon=0$ and in (b) $\epsilon=0.01$. The nointegrable term $\epsilon q^4/4$ introduces bounds to the motion and leads to the appearance of robust compact manifolds which resist to the perturbation. }
\label{fig2}
\end{figure}

Now, we consider the function
\be
\Omega(t)=1+\textrm{\scriptsize $\frac{1}{10}$}\cos t.
\ee
When $\epsilon=0$ the dynamics is marked by the so called {\em parametric resonance}~\cite{arnold}, which yield a highly unstable dynamical behavior. According to our result, in this regime the dynamics, though unstable, is integrable; no chaos is expected. Interestingly, however, the system is {\em not} Liouville integrable. The reason can be understood from the stroboscopic map of Fig.~\ref{fig2}-(a), where points $x_n$ are plotted at instants $t_n=2\pi n/\omega$, with $\omega=1$. The resonant motion, typical of forced oscillators, is such that the orbits always access farther regions as the system evolves in time. It follows that the invariant manifolds are clearly noncompact. Recurrence and transitivity do not occur and hence the motion is nonchaotic. Although integrable, the system does not fulfill the conditions for Liouville integrability. In fact, it is immediate from Fig.~\ref{fig2}-(a) that one could not find angle-action variables. Fig.~\ref{fig2}-(b), on the other hand, in contrast with Fig.~\eqref{fig1}-(b), shows an interesting situation in which the breakdown in the algebra is not followed by a clear emergence of chaos. Actually, under the introduction of the quartic term the system seems to become Liouville integrable for that value of $\epsilon$. Although this behavior is not typical---chaotic behavior is indeed observed for larger values of $\epsilon$---it illustrates the fact that Proposition 1 establishes conditions that are not {\em necessary} for integrability.

\begin{figure}
\centerline{\includegraphics[scale=0.7]{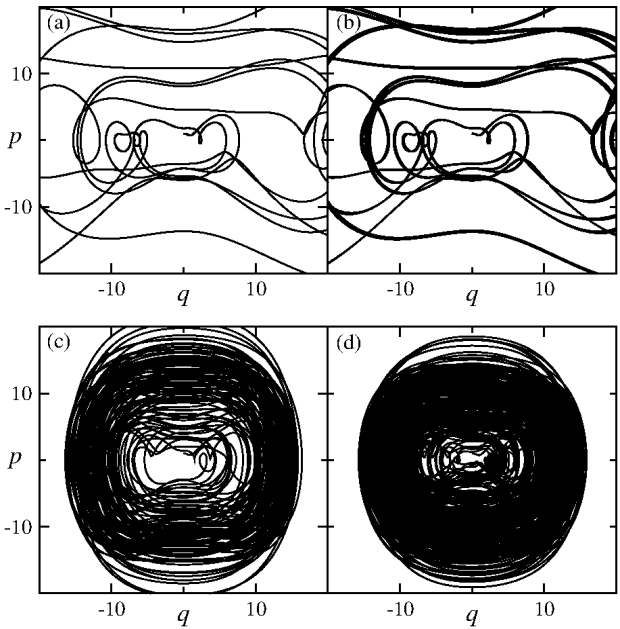}}
\caption{Time evolution of (a,c) the initial condition $x_0=(0,1)$ and (b,d) eight initial conditions located close to $x_0$ for $\Omega(t)=\frac{1}{2}\Big[\cos(e t/2)+\cos(\pi t/2)\Big]$. In (a,b) $\epsilon=0$ and in (c,d) $\epsilon=0.01$. In (b) we see that the orbits neither densely fulfill the accessible space nor departure significantly from each other. In (c) a single trajectory integrated for long times ($t=600$) fulfill a region equivalent to that occupied in (d) by eight orbits integrated for short times ($t=200$)---an expression of mixing.}
\label{fig3}
\end{figure}

\begin{figure}
\centerline{\includegraphics[scale=0.6]{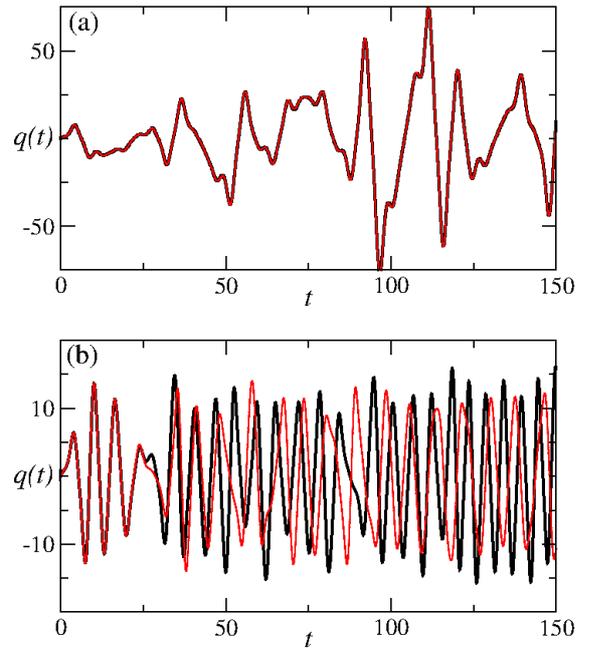}}
\caption{Time evolution of the canonical coordinate $q$ for the initial conditions $x_0=(0,1)$ (black thick line) and $x_1=(0.01,1.01)$ (red line), for $\Omega(t)=\frac{1}{2}[\cos(e t/2)+\cos(\pi t/2)]$. In (a) $\epsilon=0$ and in (b) $\epsilon=0.01$. Sensitivity to initial conditions does not occur when the algebra is finite.}
\label{fig4}
\end{figure}

Even the cases of quasiperiodic functions and nonlinear Lie algebras are shown to be correctly diagnosed by our result. First, consider the function
\be
\Omega(t)=\frac{1}{2}\Big[\cos(e t/2)+\cos(\pi t/2)\Big],
\ee 
with two manifestly incommensurable frequencies. In Fig.~\ref{fig3}, orbits are compared for finite $(\epsilon=0)$ and open $(\epsilon=0.01)$ Lie algebras. In contrast with the case of finite algebra, when $\epsilon=0.01$ a single long orbit (i.e., integrated for long times) fulfills a region equivalent to that occupied by eight short orbits (Fig.~\ref{fig3}-(c,d)). This behavior---a manifestation of mixing---is not observed when $\epsilon=0$ (Fig.~\ref{fig3}-(a,b)). In fact, this is corroborated by Fig.~\ref{fig4}, which shows the time evolution of the canonical coordinate $q$ for two close initial conditions, namely, $x_0=(0,1)$ and $x_1=(0.01,1.01)$. Clearly, sensitivity to initial condition does not occur for $\epsilon=0$. Therefore, also in the quasiperiodic case, we see that the finite algebra prevents chaotic motion to exist.

Finally, we consider the nonlinear Hamiltonian 
\be
H(x,t)=\mathbf{B}(t)\cdot\mathbf{O}(x),
\ee 
where $\mathbf{O}(x)=~(\sqrt{1-q^2}\cos p,\sqrt{1-q^2}\sin p,-q)$. For quasiperiodic magnetic fields, i.e., $\mathbf{B}=\mathbf{B}(\omega_1t,\omega_2 t)$ with incommensurable $\omega_1$ and $\omega_2$, this model was originally claimed to be chaotic~\cite{pomeau}. Soon after, however, numerical results suggested just the opposite~\cite{badii}. More recently, an argument based on Lyapunov exponents and the unitarity of the underlying quantum dynamics has been put forward attesting the integrability of the model~\cite{angelo05}. Our result supports the latter claim and yet extends it for any continuous $\mathbf{B}(t)$. Indeed, since the components of $\mathbf{O}(x)$ constitute a finite Lie algebra---actually they derive from a spin-$1/2$ Lie algebra---Proposition 1 immediately applies. 

We close this section by mentioning that it has been recently proposed, in the context of quantum mechanics, the notion of {\em upper quantum Lyapunov exponent}~\cite{jauslin04,sapin07}. By applying this quantity to the study of the quantum parametric oscillator, whose Hamiltonian is identical to that of Eq.~\eqref{H_casestudy} with $\epsilon=0$, the authors have shown to be able to distinguish between quantum regimes of stability and instability. Naturally, one may wonder whether the upper quantum Lyapunov exponent would be able to signalize the emergence of chaos under the breakdown of a Lie algebra, analogously to what we have seen in the classical case. In our case, an even more basic question arises: can one guarantee integrability, in some sense, for quantum systems linearly spanned by finite Lie algebras?

\section{Quantum integrability}

We now derive our second result and make some inferences on the integrability of a quantum analog of the Hamiltonian~\eqref{HO}. As mentioned previously, our main motivation is the manifest interest of the scientific community in quantum systems with 1.5 degrees of freedom, many of which have been claimed to display signatures of quantum chaos. In addition, we are instigated by challenging formal difficulties around the notions of integrability and chaos in quantum mechanics. In fact, the absence of a quantum analog of the Arnold-Liouville theorem seems to preclude one to rigorously formulate theses concepts. In the context delineated in this paper, however, we are led by a strong algebraic analogy with its corresponding classical system to propose a conjecture about the integrability of the underlying quantum system.

Inspired by Proposition 1, we start by proving similar results for a class of {\em quantum Lie systems}~\cite{jfc08_pla,dis,kn_10} corresponding to that defined by Eq.~\eqref{HO}.

\vskip3mm
\noindent {\bf Lemma 1.} {\em Let $\hat{H}(\hat{x},t)$ be the nonautonomous Hamiltonian of a system with a single degree of freedom, $\hat{x}=~(\hat{q},\hat{p})$, where $\hat{q}$ and $\hat{p}$ are the usual canonical operators satisfying $[\hat{q},\hat{p}]=i\hbar$. If the Hamiltonian is expansible as a linear combination of the elements $\{\hat{O}_1,\hat{O}_2,\cdots,\hat{O}_M\}$ of a finite Lie algebra, i.e.,
\begin{subequations}
\be
\hat{H}(\hat{x},t)=\mathbf{h}(t)\cdot\hat{\mathbf{O}}(\hat{x})=\sum\limits_{i=1}^{M}h_i(t)\,\hat{O}_i(\hat{x}) 
\label{HOQ}
\ee
and
\be
\frac{[\hat{O}_i,\hat{O}_j]}{i\hbar}=\sum\limits_{k=1}^M\gamma_{ij}^k\,\hat{O}_k,
\label{OcommutQ}
\ee
where $\gamma_{ij}^k=-\gamma_{ji}^k$, and if $h_i(t)$ are complex-valued functions continuous on the interval $a\leqslant t\leqslant b$, then the following holds.\\
\noindent (i) There exists a dynamical invariant of form
\be
\hat{I}(\hat{x},t)=\mathbf{g}(t)\cdot\hat{\mathbf{O}}(\hat{x})=\sum\limits_{i=1}^Mg_i(t)\,\hat{O}_i(\hat{x}),
\label{IOQ}
\ee
\end{subequations}
with $g_i(t)$ continuous on the interval. \\
\noindent (ii) The autonomous extensions $\hat{K}(\hat{x},\hat{y})$ and $\hat{I}(\hat{x},\hat{\theta})$ commute and generate linearly independent vector flows.\\
\noindent (iii) There exists a nonchaotic classical limit for $\hat{H}(\hat{x},t)$.}
\vskip3mm

The proof of Lemma 1 goes as follows.  Again, we invoke the Lewis-Reisenfeld theory~\cite{lr}, according to which an invariant of form \eqref{IOQ}, in the Schrödinger picture, is assumed to satisfy the relation
\be
\partial_t \hat{I}(\hat{x},t)+\frac{\left[ \hat{I}(\hat{x},t),\hat{H}(\hat{x},t)\right]}{i\hbar}=0,
\label{dIQ/dt}
\ee
so as to ensure that 
\be
\frac{d}{dt}\langle\hat{I}(\hat{x},t)\rangle_{\hat{\rho}}=0,
\label{d<I>/dt}
\ee 
for arbitrary $\hat{\rho}(0)$, where $\langle\hat{I}(\hat{x},t)\rangle_{\hat{\rho}}=\textrm{Tr}[\hat{\rho}(t)\hat{I}(\hat{x},t)]$. By plugging Eqs.~\eqref{HOQ} and \eqref{IOQ} into Eq.~\eqref{dIQ/dt} one obtains Eq.~\eqref{dg/dt}, which leads to the same conclusion of the classical case and proves part (i) of Lemma 1.

Let us construct the autonomous extension of $\hat{H}(\hat{x},t)$. Our problem of initial value can be written, in the Heisenberg picture, as
\be
\begin{array}{l}
\displaystyle{\frac{d \hat{x}_{\hp}}{dt}=\frac{[\hat{x}_{\hp},\hat{H}_{\hp}]}{i\hbar}}, \\ \\
\hat{\rho}(0)=\hat{\rho}_0,
\end{array} \label{HPH}
\ee
where $\hat{x}_{\hp}=\hat{U}^{\dag}(t)\,\hat{x}\,\hat{U}(t)$, $\hat{H}_{\hp}=\hat{H}(\hat{x}_{\hp},t)$  and $\hat{U}(t)$ is a unitary propagator, solution of $i\hbar\,\partial_t\hat{U}(t)=\hat{H}(\hat{x},t)\hat{U}(t)$ with $\hat{U}(0)=\mathbf{1}$. Introduce an {\em extended} dynamical problem,
\be
\begin{array}{l}
\displaystyle{\frac{d \hat{x}_{\hp}}{dt}=\frac{[\hat{x}_{\hp},\hat{K}_{\hp}]}{i\hbar}},\qquad
\displaystyle{\frac{d \hat{y}_{\hp}}{dt}=\frac{[\hat{y}_{\hp},\hat{K}_{\hp}]}{i\hbar}},\\ \\
\hat{\varrho}(0)=\hat{\rho}_0\otimes\ket{0}\bra{0},
\end{array}\label{HPK}
\ee
governed by the {\em autonomous} Hamiltonian
\be
\begin{array}{l}
\hat{K}(\hat{x},\hat{y}):= \hat{H}(\hat{x},\hat{\theta})+\hat{J},
\end{array}
\ee
with an additional degree of freedom $\hat{y}=(\hat{\theta},\hat{J})$, such that $\hat{\theta}\ket{0}=0$ and $[\hat{\theta},\hat{J}]=i\hbar$. The equivalence between the nonautonomous and autonomous dynamics is established as follows. First, note that the Heisenberg equations for $\hat{x}_{\hp}$ are decoupled from $\hat{J}_{\hp}$ and, moreover, $\hat{\theta}_{\hp}=\hat{\theta}+t$ and $\bra{0}\hat{\theta}_{\hp}\ket{0}=t$. This implies that $\hat{x}_{\hp}=\hat{x}_{\hp}(\hat{x},\hat{\theta},t)$ and $\bra{0}\hat{x}_{\hp}\ket{0}$ is the solution of the nonautonomous problem~\eqref{HPH}. Finally, in terms of the expectation value of an arbitrary operator $\hat{A}(\hat{x},\hat{\theta})$, it follows that
\be
\textrm{Tr}_{xy}\left[\hat{\varrho}(0)\,\hat{A}(\hat{x}_{\hp},\hat{\theta}_{\hp})\right]=\textrm{Tr}_{x}\left[\hat{\rho}(0)\,\hat{A}(\hat{x}_{\hp},t)\right].
\label{equivalence}
\ee
This procedure is the Heisenberg equivalent to Howland's method~\cite{howland,jauslin}. It takes us to an autonomous formulation within which the new Hamiltonian $\hat{K}$ is a constant of motion, a fact which can be proved from the Heisenberg equation for $\hat{J}_{\hp}$. 

Now we show that, under the replacement $t\to~\hat{\theta}$, $\hat{I}(\hat{x},t)$ commutes with the autonomous Hamiltonian $\hat{K}(\hat{x},\hat{y})$. From Eqs.~\eqref{equivalence} and \eqref{d<I>/dt} it follows, respectively, that $\langle \hat{I}(\hat{x},\hat{\theta})\rangle_{\hat{\varrho}}= \langle \hat{I}(\hat{x},t)\rangle_{\hat{\rho}}$ and  $\frac{d}{dt}\langle\hat{I}(\hat{x},\hat{\theta})\rangle_{\hat{\varrho}}=0$, which for an arbitrary $\hat{\rho}_0$ implies that $[\hat{I}(\hat{x},\hat{\theta}),\hat{K}(\hat{x},\hat{y})]=~0$.

The proof of flow independence follows the classical approach, with some adaptations. 

\vskip3mm\noindent
{\bf Definition 3.} {\em Let the flow direction produced by an invariant $\hat{I}_{k\hp}$ be defined by the velocity $\hat{\mathbf{v}}_{\hat{I}_{k\hp}}= (d\hat{\mathbf{r}}_{\hp}/dt)_{\hat{I}_k}=\hat{\mathcal{L}}\hat{I}_k$, where  $\hat{\mathcal{L}}\hat{I}_{k\hp}:= [\hat{\mathbf{r}}_{\hp},\hat{I}_{k\hp}]/i\hbar$ and $\hat{\mathbf{r}}_{\hp}=(\hat{x}_{\hp},\hat{y}_{\hp})$. The vector flow generated by the set $\{\hat{I}_{1\hp},\hat{I}_{2\hp},\cdots,\hat{I}_{N\hp}\}$ is independent if the vectors $\hat{\mathbf{v}}_{\hat{I}_{1\hp}},\hat{\mathbf{v}}_{\hat{I}_{2\hp}},\cdots,\hat{\mathbf{v}}_{\hat{I}_{N\hp}}$ are linearly independent.}
\vskip3mm

Note that $\hat{K}(\hat{x},\hat{y})=\hat{\mathbf{h}}(\hat{\theta})\cdot\hat{\mathbf{O}}(\hat{x})+\hat{J}$ and $\hat{I}(\hat{x},\hat{\theta})=\hat{\mathbf{g}}(\hat{\theta})\cdot\hat{\mathbf{O}}(\hat{x})$. Let $\hat{\mathbf{V}}_{\hp}=a_K\hat{\mathbf{v}}_{\hat{K}_{\hp}}+a_I\hat{\mathbf{v}}_{\hat{I}_{\hp}}$ be a linear combination of the velocities. It follows that
\be
\hat{\mathbf{V}}_{\hp}&=&(\hat{\mathcal{L}}_{\hat{x}_{\hp}}+\hat{\mathcal{L}}_{\hat{y}_{\hp}})(a_K\,\hat{K}_{\hp}+a_I\,\hat{I}_{\hp}) \nonumber \\
&=&(a_K\,\hat{\mathbf{h}}_{\hp}+a_I\,\hat{\mathbf{g}}_{\hp})\cdot\hat{\mathcal{L}}_{\hat{x}_{\hp}}\hat{\mathbf{O}}_{\hp}\nonumber \\ &-&\left[\partial_{\hat{\theta}_{\hp}}(a_K\,\hat{\mathbf{h}}_{\hp}+a_I\,\hat{\mathbf{g}}_{\hp})\cdot\hat{\mathbf{O}}_{\hp}\right]\mathbf{e}_{\textrm{\tiny $J$}} +a_K\,\mathbf{e}_{\textrm{\tiny $\theta$}},
\label{V_H}
\ee
where $a_{K,I}$ are arbitrary real numbers and $\mathbf{e}_{\textrm{\tiny $\theta,J$}}$ are unitary vectors denoting orthogonal directions in phase space. Clearly, as in the classical context, the only solution allowed for $\hat{\mathbf{V}}_{\hp}=0$ is $a_K=a_I=0$, which implies the linear independence between $\hat{\mathbf{v}}_{\hat{K}_{\hp}}$ and $\hat{\mathbf{v}}_{\hat{I}_{\hp}}$. Furthermore, since this conclusion is obtained at the level of operators it is automatically guaranteed that $\bra{\mu}\hat{\mathbf{v}}_{\hat{K}_{\hp}}\ket{\nu}$ and $\bra{\mu}\hat{\mathbf{v}}_{\hat{I}_{\hp}}\ket{\nu}$ are linearly independent, for whatever states $\ket{\mu}$ and $\ket{\nu}$. This completes the proof of part (ii) of Lemma~1.

Finally, it is straightforward to show that there exists a classical limit for the quantum model~\eqref{HOQ}-\eqref{IOQ} and that this classical limit precisely corresponds to the structure given by Eqs.~\eqref{HO}-\eqref{IO}, which is proven integrable. The approach consists in employing the Weyl-Wigner formalism \cite{weyl27,wigner32,polko10} to project operators in phase space and then taking the limit $\hbar\to 0$. The classical counterpart $\mathcal{A}$ of a Heisenberg operator $\hat{A}_{\hp}$ is defined here as
\be
\mathcal{A}(x,t)=W_0\{\hat{A}_{\hp}\}:=\lim\limits_{\hbar\to 0}W\{\hat{A}_{\hp}\},
\label{Acl}
\ee
where 
\be
W\{\hat{A}_{\hp}\}=\int\limits_{-\infty}^{\infty} du\,e^{\imath u p/\hbar}\langle q-u/2| \hat{A}_{\hp} |q+u/2\rangle.\qquad
\ee
The Weyl-Wigner transform, denoted by $W\{\,\cdot\,\}$, maps the quantum commutator $[\hat{A}_{\hp},\hat{B}_{\hp}]/i\hbar$ onto the Moyal bracket~\cite{moyal49}, which in the strict limit of $\hbar\to~0$ reduces to the Poisson bracket $\{\mathcal{A}(x,t),\mathcal{B}(x,t)\}$. It follows that the Heisenberg equations are mapped onto Hamilton's equations and the classical trajectory in phase space is obtained via $x=W_0\{\hat{x}_{\hp}\}$. Given these relations, it is just an exercise to show that $W_0$ maps Eqs.~\eqref{HOQ}-\eqref{IOQ} onto Eqs.~\eqref{HO}-\eqref{IO}.  \hfill{$\blacksquare$}

Although we have not provided physical meaning for the notions of involution and flow independence other than the analogy with their classical counterparts, the lemma we have just proved does identify a class of systems which satisfies conditions conventionally required---though not rigorously formulated---for the system to be classified as quantum integrable. In particular, Lemma 1 guarantees the existence of a nonchaotic classical limit. Now, even though one may argument that not every quantum system possesses a well-defined classical limit, so that this would not be a good criterium for quantum integrability, it is still reasonable to classify as quantum integrable those which do possess a proven nonchaotic classical limit. Actually, it would be rather counterintuitive to classify as quantum chaotic a system whose classical limit is proven nonchaotic.

Additionally, we should remark that our quantum model can be said integrable also in a further sense. Hamiltonians of type \eqref{HOQ} have shown to be {\em explicitly solvable} in several contexts~\cite{kn_10,lai96}. More importantly, according to a seminal result by Zhang, Feng, and Gilmore~\cite{zhangRMP}, the time-dependent Schrödinger equation for these quantum Lie Hamiltonians can be exactly solved provided the initial state is an arbitrary coherent state. It immediately follows that the Hamiltonian \eqref{HOQ} is quantum integrable---in the sense of being explicitly solvable---for the emblematic class of arbitrary minimum-uncertainty states.

Therefore, the scenario is such that the quantum model \eqref{HOQ} satisfies several criteria of quantum integrability. In fact, the existence of a nontrivial quantum invariant, the algebraic symmetry in correspondence with the classical structure, the existence of a nonchaotic classical limit, and a demonstrated explicit solvability in many contexts, define a set of properties which strongly suggests that the quantum structure has no ancestral reason to mimic chaos. In spite of these remarkable evidences, we do not have rigorous elements to ensure quantum integrability within broader frameworks~\cite{caux10}. These observations set the grounds for the following statement.

\vskip3mm
\noindent
{\bf Conjecture.} {\em A quantum system described by a Hamiltonian of form \eqref{HOQ}, spanned by a finite Lie algebra such as \eqref{OcommutQ}, is quantum integrable.}
\vskip3mm

Supported by the above arguments, this conjecture asserts that a quantum system such as \eqref{HOQ} should be diagnosed as quantum integrable in any claimed-general theory of quantum integrability. In particular, our conjecture precludes these systems to present any symptoms of quantum chaos. Whatever the precise substance this latter term may assume in each context, it often meets tools for its diagnose. Thus, we do not expect Hamiltonians of type \eqref{HOQ} to exhibit any of the well-known signatures of quantum chaos, as for instance level repulsion and Wigner level statistics (within the random matrix approach~\cite{weid09}) or Lyapunov regime (in the Loschmidt echo decay~\cite{gorin06}). These theories turn out to be, therefore, arenas for preliminary tests of our conjecture.


\subsection{Applications}

Note that our conjecture readily applies to the rather important class of quadratic systems obeying the algebra $\{\mathbf{1},\hat{q}^2,\hat{p}^2,\hat{q}\hat{p}+\hat{p}\,\hat{q},\hat{q},\hat{p}\}$, whose physical applications range from Hamiltonian cosmology to Bose-Einstein condensation~\cite{soto10,hensinger}. For this algebra, dynamical invariants and explicit solutions for the quantum dynamics are known for a variety of time-dependent parameters.

Another important class of models to which our conjecture applies is the one describing a spin-$S$ dynamics under time-dependent magnetic fields. Given a Hamiltonian $\hat{H}(t)=\mathbf{B}(t)\cdot\hat{\mathbf{S}}$, with $\hat{\mathbf{S}}=(\hat{S}_x,\hat{S}_y,\hat{S}_z)$ and $[\hat{S}_x,\hat{S}_y]=i\hbar\hat{S}_z$ plus cyclic permutations, one may employ the Holstein-Primakoff transformation,
\be 
\begin{array}{l}
\hat{S}_+=\hbar\,\hat{a}^{\dag}(2 S-\hat{a}^{\dag}\hat{a})^{1/2},\qquad \hat{S}_z=\hbar\,(S-\hat{a}^{\dag}\hat{a}),
\end{array}
\ee
where $\hat{S}_{\pm}=\hat{S}_x\pm i\hat{S}_y$, to proceed the bosonization of the system. Then, using the ordinary parametrization $\hat{a}=\frac{1}{\sqrt{2}}\left(\frac{\hat{q}}{b}+i\frac{\hat{p}}{c} \right)$, with $bc=\hbar$, one maps the original problem onto a one-dimensional Hamiltonian of form \eqref{HOQ} for which the Lie algebra $\{\hat{S}_x(\hat{x}),\hat{S}_y(\hat{x}),\hat{S}_z(\hat{x})\}$, though composed of elements with a far nontrivial dependence on $\hat{x}$, is still a closed spin algebra.

\section{Conclusion}

In summary, we have shown that time-dependent Hamiltonian systems with one degree of freedom linearly spanned by a finite Lie algebra are regular (nonchaotic) and mostly Liouville integrable. Differently from many traditional approaches which look for the explicit solvability of Lie systems, we have used the algebraic features of the model to prove, in the lines of the Arnold-Liouville theorem, the absence of chaos. Our result settles the polemic around the Rabi oscillations in two-level systems under generic magnetic fields and gives elements to address many problems involving quasiperiodic fields.

In direct analogy with the classical context, we have stated a conjecture about the integrability of quantum systems with analog algebraic features. Our proposal, which claims the nonexistence of quantum chaos for these systems, is not only based on the existence of a well-defined classical limit, proven nonchaotic, but also on a strong Lie-algebraic correspondence between the quantum and classical structures as well as on the demonstrated explicit solvability of the model in several contexts. Still, we have suggested the notion of quantum flow independence, which is manifestly inspired by its classical counterpart and may hopefully be used in further elaborations on the notion of quantum integrability.

\section*{Acknowledgments}
A.D.R. and R.M.A. acknowledge financial support from CNPq, and E.I.D. from FAPEMIG. The authors would like to thank L. A. R. de Santana, L. Sanz, W. F. Wreszinski, and M. A. M. de Aguiar for helpful discussions. We are also indebited to F. Hass for valuable comments on an early version of this manuscript.



\begin{thebibliography}{99}

\bibitem{gutz}  M. C. Gutzwiller, {\em Chaos in Classical and Quantum Mechanics} (Springer-Verlag, New York, 1990).

\bibitem{ozorio} 
A. M. Ozorio de Almeida, {\em Hamiltonian Systems: Chaos and Quantization} (Cambridge University Press, Cambridge, 1988).

\bibitem{arnold} V. I. Arnold, {\em Mathematical Methods of Classical Mechanics} (Springer-Verlag, New York, 1978).

\bibitem{pomeau} Y. Pomeau, B. Dorizzi and B. Grammaticos, Phys. Rev. Lett. {\bf 56}, 681 (1986).

\bibitem{hensinger} W. K. Hensinger {\em et al}, Nature {\bf 412}, 52 (2001).

\bibitem{soto10} R. Cordero-Soto, E. Suazo, and S. K. Suslov, Ann. Phys. (N.Y.) {\bf 325}, 1884 (2010).

\bibitem{shudo} A. Ishikawa, A. Tanaka, and Akira Shudo, Phys. Rev. Lett. {\bf 104}, 224102 (2010).

\bibitem{badii} R. Badii and P. F. Meier, Phys. Rev. Lett. {\bf 58}, 1045 (1987).

\bibitem{angelo05} R. M. Angelo and W. F. Wreszinski, Phys. Rev. A {\bf 72}, 034105 (2005); Ann. Phys. (N.Y.) {\bf 322}, 769 (2007).

\bibitem{jfc00} J. F. Cariñena and J. Nasarre, J. Opt. B: Q. Semiclass. Opt. {\bf 2}, 94, (2000).

\bibitem{jfc08_jpa} J. F. Cariñena, J. de Lucas, and M. F. Rañada,  J. Phys. A {\bf 41} 304029 (2008).

\bibitem{jfc08_pla} J. F. Cariñena and J. de Lucas, Phys. Lett. A {\bf 372}, 5385 (2008); Int. J. Geom. Meth. Mod. Phys. {\bf 6}, 1235 (2009).

\bibitem{dis} J. F. Cariñena and J. de Lucas, {\em Lie systems: theory, generalisations, and applications}, Dissertationes mathematicae {\bf 479} (Warszsawa, 2011); arXiv:1103.4166. 

\bibitem{rfe_11} R. Flores-Espinoza, Int. J. Geom. Meth. Modern Phys. {\bf 8}, 1169 (2011).

\bibitem{kn_10} M. Kuna and J. Naudts, Rep. Math. Phys. {\bf 65}, 77 (2010).

\bibitem{korsch82} H. J. Korsch, Phys. Lett. A {\bf 90}, 113 (1982).

\bibitem{hietarinta82} J. Hietarinta, Phys. Lett. A {\bf 93}, 55 (1982).

\bibitem{hietarinta84} J. Hietarinta, J. Math. Phys. {\bf 25}, 1833 (1984).

\bibitem{zhang89} W. M. Zhang, D. H. Feng, J. M. Yuan, and S. J. Wang, Phys. Rev. A {\bf 40}, 438 (1989).

\bibitem{weigert92} S. Weigert, Physica D {\bf 56}, 107 (1992).

\bibitem{gallardo09} J. C. Gallardo and G. Marmo, Int. J. Geom. Met. Mod. Phys. {\bf 6}, 129 (2009).

\bibitem{zhang90} W. M. Zhang, D. H. Feng, and J. M. Yuan, Phys. Rev. A {\bf 42}, 7125 (1990).

\bibitem{caux10} J. S. Caux and J. Mossel, J. Stat. Mech. P02023 (2011); arXiv:1012.3587v1 (2010).

\bibitem{weid09} H. A. Weidenmüller and G. E. Mitchell, Rev. Mod. Phys. {\bf 81}, 539 (2009).

\bibitem{gorin06} T. Gorin, T. Prosen, T. H. Seligman, and M. \v{Z}nidari\v{c}, Phys. Reports {\bf 435}, 33 (2006).

\bibitem{noether} E. Noether, Nachr. Ges. Wiss. Goettingen {\bf 57}, 235 (1918).

\bibitem{lutzky} M. Lutzky, Phys. Lett. A {\bf 68}, 3 (1978).

\bibitem{lewis} H. R. Lewis Jr., Phys. Rev. Lett. {\bf 18}, 510 (1967); J. Math. Phys. {\bf 9}, 1976 (1968).

\bibitem{korsch79} H. J. Korsch, Phys. Lett. A {\bf 74}, 294 (1979).

\bibitem{kaushal81} R. S. Kaushal and H. J. Korsch, J. Math. Phys. {\bf 22}, 1904 (1981).

\bibitem{kaushal93} R. S. Kaushal and S. C. Mishra, J. Math. Phys. {\bf 34}, 5843 (1993).

\bibitem{bouquet98} S. Bouquet and A. Bourdier, Phys. Rev. E {\bf 57}, 1273 (1998).

\bibitem{riedel01} J. Struckmeier and C. Riedel, Phys. Rev. E {\bf 64}, 026503 (2001).

\bibitem{vazquez08} J. L. Fu, S. Jiménez, Y. F. Tang, and L. Vázquez, Phys. Lett. A {\bf 372}, 1555 (2008).

\bibitem{yeh93} L. Yeh, Phys. Rev. A {\bf 47}, 3587 (1993).

\bibitem{nunez94} D. B. Monteoliva, H. J. Korsch, and J. A. Núñez, J. Phys. A {\bf 27}, 6897 (1994).

\bibitem{bandy01} J. N. Bandyopadhyay, A. Lakshminarayan, and V. B. Sheorey, Phys. Rev. A {\bf 63}, 042109 (2001).

\bibitem{shen03} J. Q. Shen, H. Y. Zhu, and P. Chen, Eur. Phys. J. D {\bf 23}, 305 (2003).

\bibitem{lima08} A. L. Lima, A. Rosas, and I. A. Pedrosa, Ann. Phys. (N.Y.) {\bf 323}, 2253 (2008).

\bibitem{suslov10} S. K. Suslov, Phys. Scr. {\bf 81}, 055006 (2010).

\bibitem{lr} H. R. Lewis Jr. and W. B. Riesenfeld, J. Math. Phys. {\bf 10}, 1458 (1969).

\bibitem{howland} J. S. Howland, Lecture Notes in Physics {\bf 130}, 163 (1980).

\bibitem{jauslin} H. R. Jauslin, {\em Stability and Chaos in Classical and Quantum Hamiltonian Systems}, in: P. L. Garrido, J. Marro (Eds.), II Granada Lectures on Computational Physics, 107 (World Scientific, Singapore, 1993).

\bibitem{lich} A. J. Lichtenberg and M. A. Lieberman, {\em Regular and Chaotic Dynamics}, 2nd edition  (Springer-Verlag, New York, 1992).

\bibitem{weyl27} H. Weyl, Z. Phys. {\bf 46}, 1 (1927).

\bibitem{wigner32} E. Wigner, Phys. Rev. {\bf 40}, 749 (1932).

\bibitem{polko10} A. Polkovnikov, Ann. Phys. (N.Y.) {\bf 325}, 1790 (2010).

\bibitem{moyal49} J. E. Moyal, Mathematical Proceedings of the Cambridge Philosophical Society {\bf 45}, 99 (1949).

\bibitem{devaney} R. L. Devaney, {\em An Introduction to Chaotic Dynamical Systems}, 2nd edition (Addison-Wesley Publishing Company, Reading, Massachusetts, 1989).

\bibitem{hasselblatt} B. Hasselblatt and A. Katok, {\em A First Course in Dynamics: With a Panorama of Recent Developments} (Cambridge University Press, Cambridge (UK), 2003).

\bibitem{jose} J. V. José and E. J. Saletan, {\em Classical Dynamics: A Contemporary Approach} (Cambridge University Press, Cambridge (UK), 1998).

\bibitem{hartman} P. Hartman, {\em Ordinary differential equations} (Sian, Philadelphia, 2002). 

\bibitem{jauslin04} H. R. Jauslin, O. Sapin, S. Guérin, and W. F. Wreszinski, J. Math. Phys. {\bf 45}, 4377 (2004).

\bibitem{sapin07} O. Sapin, H. R. Jauslin, and S. Weigert, J. Stat. Phys. {\bf 4}, 699 (2007).

\bibitem{lai96} Y.-Z. Lai, J.-Q. Liang, H. J. W. Müller-Kirsten, and J.-G. Zhou, Phys. Rev. A {\bf 53}, 3691 (1996).

\bibitem{zhangRMP} W.-M. Zhang, D. H. Feng, and R. Gilmore, Rev. Mod. Phys. {\bf 62}, 867 (1990).



\end{thebibliography}
\end{document}